\documentclass[prd,twocolumn,showpacs,floatfix,amsmath,amssymb,floatfix]{revtex4}
\usepackage{graphicx,color,dcolumn,booktabs,bm}
\usepackage{longtable,lscape}
\usepackage{txfonts}
\usepackage{overpic}
\usepackage{amssymb}
\usepackage{indentfirst}
\usepackage{feynmf}   
\usepackage{slashed}  
\usepackage{cases}
\usepackage{color}
\usepackage{multirow}
\usepackage{graphicx,color,dcolumn,booktabs,bm}

\graphicspath{{Figures/}} %

\begin{document}

\title{Exploration of charmed pentaquarks}
\author{S. M. Gerasyuta}\email{gerasyuta@sg6488.spb.edu}
\author{V. I. Kochkin}\email{vik@efa.ru}
\affiliation{Department of Physics, St. Petersburg State Forest Technical University, Institutski Per. 5, St. Petersburg 194021, Russia}
\author{Xiang Liu$^{1,2}$}\email{xiangliu@lzu.edu.cn}
\affiliation{$^1$Research Center for Hadron and CSR Physics,
Lanzhou University $\&$ Institute of Modern Physics of CAS,
Lanzhou 730000, China\\
$^2$School of Physical Science and Technology, Lanzhou University,
Lanzhou 730000, China}

\begin{abstract}
In this work, we explore the charmed pentaquarks, where the relativistic five-quark equations are obtained by the dispersion relation technique.
By solving these equations with the method based on the extraction of the leading singularities of the amplitudes, we predict the mass spectrum of
charmed pentaquarks with $J^P=1/2^{\pm}$ and $3/2^{\pm}$, which is valuable to further experimental study of charmed pentaquark.
\end{abstract}

 \maketitle

\section{introduction}\label{sec1}

Exploring and investigating exotic states, which include glueball, hybrid state and multiquark states, are an intriguing research topic in particle physics.
With more and more observations of new hadronic states, there were extensive discussions of whether these observed new hadronic states are good candidates
of exotic states (see Refs. \cite{1, 2} for a recent review).  Studying the hadronic configuration beyond the conventional meson and
baryon can make our knowledge of
non-perturbative QCD be abundant.

In 2013, the BESIII Collaboration announced the observation of the charged charmonium-like structure $Z_c(3900)$ in the $J/\psi^\pm$ invariant mass
spectrum of $e^+e^-\to J/\psi\pi^+\pi^-$ at $\sqrt{s}=4.26$ GeV \cite{3}. $Z_c(3900)$ can be a good candidate of the $D\bar{D}^*$ molecular
state \cite{4, 5}, which is as one of the four-quark matters. If four-quark matter is possible existing in nature, we naturally conjecture
whether there exist pentaquark  states.

In 2003,  the $\gamma^{12}C\to K^+K^-X$ reaction was studied and a peak was found in the $K^+n$ invariant mass spectrum around 1540 MeV, which was identified
as a signal for a pentaquark with positive strangeness, the ``$\Theta^+(1540)$'' \cite{6} . The unexpected finding lead to a large number of poor
statistic experiments where a positive signal was also found, but gradually an equally big number of large statistic experiments showed no evidence for such
a peak. A comprehensive review of these developments was done in \cite{7}, where one can see the relevant literature on the subject, as well as in
the devoted section of Particle Data Group (PDG) \cite{8}.

Although the signal of $\Theta^+(1540)$ was not confirmed in experiment, searching for pentaquark is still an important task \cite{9}. Thus, we need
to carry out further theoretical study of pentaquark, which can provide us more abundant information of possible pentaquark. We also notice that most of new
hadronic states were observed in the charm-$\tau$ energy region. This fact shows that the charm-$\tau$ energy region should be a suitable platform to study
pentaquark. Especially, the $Z_c(3900)$ observation boosts our confidence to study heavy flavor pentaqurk again.

In this work, we focus on the charmed pentaquark states with $J^P=1/2^\pm,3/2^\pm$, which are composed of a charm antiquark and four light quarks. Firstly,
we need to construct relativistic five-quark equations, which contain the $u$, $d$, and $c$ quarks. And then, the masses of these discussed pentaquarks can be
determined by the poles of these amplitudes, where the constituent quark involved in our calculation is the color triplet and the quark amplitudes obey the
global color symmetry. As the main task of this work, we need to perform the calculation of the pentaquark amplitudes which contain the contribution of four
subamplitudes: molecular subamplitude $BM$, $D\bar{q}D$, $Mqqq$ subamplitudes and $Dqq\bar{q}$ subamplitude ($D$ denotes the diquark state, $B$ and $M$ are
the baryon and meson states), where the relativistic generalization of five-quark Faddeev-Yakubovsky equations is constructed in the form of the dispersion
relation \cite{10}.
Finally, we can get the masses of the low-lying charmed pentaquarks, which provide valuable information to further experimental search for these predicted
charmed pentaquarks.

Our paper is organized as follows. After this introduction, propose the brief discussion of relativistic Faddeev equations. In Sec. \ref{sec3}
we represent the five-quark amplitudes relevant to charmed pentaquarks. The numerical results are shown in Sec. \ref{sec4}. The last section is
devoted to a summary.

\section{Brief introduction of relativistic Faddeev equations}\label{sec2}

We consider the derivation of the relativistic generalization of the
Faddeev equation for the example of the $\Delta$-isobar
($J^P=\frac{3}{2}^+$). This is convenient because the spin-flavour part
of the wave function of the $\Delta$-isobar contains only nonstrange quarks
and pair interactions with the quantum numbers of a $J^P=1^+$ diquark
(in the color state $\bar 3_c$). The $3q$ baryon state $\Delta$ is
constructed as color singlet. Suppose that there is a $\Delta$-isobar
current which produces three $u$ quarks (Fig. 1a). Successive pair
interactions lead to the diagrams shown in Fig. 1b-1f. These diagrams can
be grouped according to which of the three quark pairs undergoes the last
interaction i.e., the total amplitude can be represented as a sum of
diagrams. Taking into account the equality of all pair interactions of
nonstrange quarks in the state with $J^P=1^+$, we obtain the corresponding
equation for the amplitudes

\begin{equation}
A_1 (s, s_{12}, s_{13}, s_{23})=\lambda+A_1 (s, s_{12})+
A_1 (s, s_{13})+A_1 (s, s_{23})\, . \end{equation}

\noindent
Here the $s_{ik}$ are the pair energies of particles 1, 2 and 3, and $s$
is the total energy of the system. Using the diagrams of  Fig. 1, it is
easy to write down a graphical equation for the function $A_1 (s, s_{12})$
(Fig. 2). To write down a concrete equations for the function
$A_1 (s, s_{12})$ we must specify the amplitude of the pair interaction
of the quarks. We write the amplitude of the interaction of two quarks in
the state $J^P=1^+$ in the form:

\begin{equation}
a_1(s_{12})=\frac{G^2_1(s_{12})}
{1-B_1(s_{12})} \, ,\end{equation}

\begin{equation}
B_1(s_{12})=\int\limits_{4m^2}^{\infty}
\, \frac{ds'_{12}}{\pi}\frac{\rho_1(s'_{12})G^2_1(s'_{12})}
{s'_{12}-s_{12}} \, ,\end{equation}

\begin{eqnarray}
\rho_1 (s_{12})&=&
\left(\frac{1}{3}\, \frac{s_{12}}{4m^2}+\frac{1}{6}\right)
\left(\frac{s_{12}-4m^2}{s_{12}}\right)^{\frac{1}{2}} \, .
\end{eqnarray}

\noindent
Here $G_1(s_{12})$ is the vertex function of a diquark with $J^P=1^+$.
$B_1(s_{12})$ is the Chew-Mandelstam function \cite{11} and $\rho_1 (s_{12})$
is the phase spaces for a diquark with $J^P=1^+$.

The pair quarks amplitudes $qq \to qq$ are calculated in the framework of
the dispersion $N/D$ method with the input four-fermion interaction
\cite{12, 13} with the quantum numbers of the gluon \cite{14, 15}.

The four-quark interaction is considered as an input:

\begin{eqnarray}
 & g_V \left(\bar q \lambda I_f \gamma_{\mu} q \right)^2 +
2\, g^{(s)}_V \left(\bar q \lambda I_f \gamma_{\mu} q \right)
\left(\bar s \lambda \gamma_{\mu} s \right)+
g^{(ss)}_V \left(\bar s \lambda \gamma_{\mu} s \right)^2
 \, . & \end{eqnarray}

\noindent
Here $I_f$ is the unity matrix in the flavour space $(u, d)$. $\lambda$ are
the color Gell-Mann matrices. Dimensional constants of the four-fermion
interaction $g_V$, $g^{(s)}_V$ and $g^{(ss)}_V$ are parameters of the
model. At $g_V =g^{(s)}_V =g^{(ss)}_V$ the flavour $SU(3)_f$ symmetry occurs.
The strange quark violates the flavour $SU(3)_f$ symmetry. In order to
avoid additional violation parameters we introduce the scale of the
dimensional parameters \cite{15}:

\begin{equation}
g=\frac{m^2}{\pi^2}g_V =\frac{(m+m_s)^2}{4\pi^2}g_V^{(s)} =
\frac{m_s^2}{\pi^2}g_V^{(ss)}
\, ,\end{equation}

\begin{eqnarray}
\nonumber
\Lambda=\frac{4\Lambda(ik)}
{(m_i+m_k)^2}. \nonumber \end{eqnarray}

\noindent
Here $m_i$ and $m_k$ are the quark masses in the intermediate state of
the quark loop. Dimensionless parameters $g$ and $\Lambda$ are supposed
to be constants which are independent of the quark interaction type. The
applicability of Eq. (5) is verified by the success of
De Rujula-Georgi-Glashow quark model \cite{14}, where only the short-range
part of Breit potential connected with the gluon exchange is responsible
for the mass splitting in hadron multiplets.

In the case under discussion the interacting pairs of particles do not
form bound states. Therefore, the integration in the dispersion integral (7)
run from $4m^2$ to $\infty$. The equation corresponding to Fig. 2 can be
written in the form:

\begin{eqnarray}
A_1(s,s_{12})&
=&\frac{\lambda_1 B_1(s_{12})}{1-B_1(s_{12})}\nonumber\\
&&\nonumber\\
&+&\frac{G_1(s_{12})}{1-B_1(s_{12})}
\int\limits_{4m^2}^{\infty}
\, \frac{ds'_{12}}{\pi}\frac{\rho_1(s'_{12})}
{s'_{12}-s_{12}}G_1(s'_{12})\nonumber\\
&&\nonumber\\
 & \times & \int\limits_{-1}^{+1} \, \frac{dz}{2}
[A_1(s,s'_{13})+A_1(s,s'_{23})] \, .
\end{eqnarray}

In Eq. (7) $z$ is the cosine of the angle between the relative momentum
of particles 1 and 2 in the intermediate state and the momentum of the third
particle in the final state in the c.m.s. of the particles 1 and 2. In
our case of equal mass of the quarks 1, 2 and 3, $s'_{13}$ and $s'_{12}$
are related by Eq. (8) (See Ref. \cite{16})

\begin{eqnarray}
s'_{13}&=&2m^2-\frac{(s'_{12}+m^2-s)}{2}\\
&&\nonumber\\
\nonumber
 & \pm & \frac{z}{2} \sqrt{\frac{(s'_{12}-4m^2)}{s'_{12}}
(s'_{12}-(\sqrt{s}+m)^2)(s'_{12}-(\sqrt{s}-m)^2)}\, .
\end{eqnarray}

The expression for $s'_{23}$ is similar to (8) with the replacement
$z\to -z$. This makes it possible to replace
$[A_1(s,s'_{13})+A_1(s,s'_{23})]$ in (7) by $2A_1(s,s'_{13})$.

From the amplitude $A_1(s,s_{12})$ we shall extract the singularities
of the diquark amplitude:

\begin{eqnarray}
A_1(s,s_{12})=
\frac{\alpha_1(s,s_{12}) B_1(s_{12})}{1-B_1(s_{12})} \, .
\end{eqnarray}

The equation for the reduced amplitude $\alpha_1(s,s_{12})$ can be written as

\begin{eqnarray}
\alpha_1(s,s_{12})&=&\lambda+\frac{1}{B_1(s_{12})}
\int\limits_{4m^2}^{\infty}
\, \frac{ds'_{12}}{\pi}\frac{\rho_1(s'_{12})}
{s'_{12}-s_{12}}G_1(s'_{12})
\nonumber\\
&&\nonumber\\
& \times &
\int\limits_{-1}^{+1} \, \frac{dz}{2}
\, \frac{2\alpha_1(s,s'_{13}) B_1(s'_{13})}{1-B_1(s'_{13})} \, .
\end{eqnarray}

The next step is to include into (10) a cutoff at large $s'_{12}$. This
cutoff is needed to approximate the contribution of the interaction at
short distances. In this connection we shall rewrite Eq. (10) as

\begin{eqnarray}
\alpha_1(s,s_{12})&=&\lambda+\frac{1}{B_1(s_{12})}
\int\limits_{4m^2}^{\infty}
\, \frac{ds'_{12}}{\pi}\Theta(\Lambda-s'_{12})\frac{\rho_1(s'_{12})}
{s'_{12}-s_{12}}G_1
\nonumber\\
&&\nonumber\\
& \times &
\int\limits_{-1}^{+1} \, \frac{dz}{2}
\, \frac{2\alpha_1(s,s'_{13}) B_1(s'_{13})}{1-B_1(s'_{13})} \, .
\end{eqnarray}

In Eq. (11) we have chosen a hard cutoff. However, we can also use a soft
cutoff, for instance
$G_1(s'_{12})=G_1 \exp \left( -\frac{(s'_{12}-4m^2)^2}{\Lambda^2} \right)$,
which leaves the results of calculations of the mass spectrum essentially
unchanged.

The construction of the approximate solution of Eq. (11) is based on
extraction of the leading singularities are close to the region
$s_{ik}\approx 4m^2$. The structure of the singularities of amplitudes
with a different number of rescattering (Fig. 1) is the following \cite{16}.
The strongest singularities in $s_{ik}$ arise from pair rescatterings of
quarks: square-root singularity corresponding to a threshold and pole
singularities corresponding to bound states (on the first sheet in the case
of real bound states, and on the second sheet in the case of virtual bound
states). The diagrams of Figs. 1b and 1c have only these two-particle
singularities. In addition to two-particle singularities diagrams of
Figs. 1d and 1e have their own specific triangle singularities. The
diagram of Figs. 1f describes a larger number of three-particle
singularities. In addition to singularities of triangle type it contains
other weaker singularities. Such a classification of singularities
makes it possible to search for an approximate solution of Eq. (11),
taking into account a definite number of leading singularities and
neglecting the weaker ones. We use the approximation in which the
singularity corresponding to a single interaction of all three particles,
the triangle singularity, is taken into account.

For fixed values of $s$ and $s'_{12}$ the integration is carried out
over the region of the variable $s'_{13}$ corresponding to a physical
transition of the current into three quarks (the physical region of
Dalitz plot). It is convenient to take the central point of this region,
corresponding to $z=0$, to determinate the function $\alpha_1(s,s_{12})$
and also the Chew-Mandelstam function $B_1(s_{12})$ at the point
$s_{12}=s_0=\frac{s}{3}+m^2$. Then the equation for the $\Delta$ isobar
takes the form:

\begin{equation}
\alpha_1(s,s_0)=\lambda+I_{1,1}(s,s_0)\cdot 2\, \alpha_1(s,s_0)
\, , \end{equation}

\begin{equation}
I_{1,1}(s,s_0)=\int\limits_{4m^2}^{\Lambda_1}
\, \frac{ds'_{12}}{\pi} \frac{\rho_1(s'_{12})}
{s'_{12}-s_{12}}G_1\int\limits_{-1}^{+1} \, \frac{dz}{2}
\, \frac{G_1}{1-B_1(s'_{13})}
\, . \end{equation}

We can obtain an approximate solution of Eq. (12)

\begin{equation}
\alpha_1(s,s_0)=\lambda [1-2\, I_{1,1}(s,s_0)]^{-1}
\, . \end{equation}

The function $I_{1,1}(s,s_0)$ takes into account correctly the singularities
corresponding to the fact that all propagators of triangle diagrams like
those of Figs. 1d and 1e reduce to zero. The right-hand side of Eq. (14) may
have a pole in $s$, which corresponds to a bound state of the three quarks.
The choice of the cutoff $\Lambda$ makes it possible to fix the value of
the mass of the $\Delta$ isobar.

Baryons of $S$-wave multiplets have a completely symmetric spin-flavour
part of the wave function, and spin $\frac{3}{2}$ corresponds to the
decuplet which has a symmetric flavour part of the wave function. Octet
states have spin $\frac{1}{2}$ and a mixed symmetry of the flavour function.

In analogy with the case of the $\Delta$ isobar we can obtain the
rescattering amplitudes for all $S$-wave states with $J^P=\frac{3}{2}^+$,
which include quarks of various flavours. These amplitudes will satisfy
systems of integral equations. In considering the $J^P=\frac{1}{2}^+$
octet we must include the interaction of the quarks in the $0^+$ and $1^+$
states (in the colour state $\bar 3_c$). Including all possible rescattering
of each pair of quarks and grouping the terms according to the final states
of the particles, we obtain the amplitudes $A_0$ and $A_1$, which satisfy
the corresponding systems of integral equations. If we choose the
approximation in which two-particle and triangle singularities are taken
into account, and if all functions which depend on the physical region of
the Dalitz plot, the problem of solving the system of integral equations
reduces to one of solving simple algebraic equations.

In our calculation the quark masses $m_u=m_d=m$ and $m_s$ are not uniquely
determined. In order to fix $m$ and $m_s$ anyhow, we make the simple
assumption that $m=\frac{1}{3} m_{\Delta} (1232)$
$m=\frac{1}{3} m_{\Omega} (1672)$. The strange quark breaks the flavour
$SU(3)_f$ symmetry (6).

In Ref. \cite{17} we consider two versions of calculations. If the first version
the $SU(3)_f$ symmetry is broken by the scale shift of the dimensional
parameters. A single cutoff parameter in pair energy is introduced for all
diquark states $\lambda_1=12.2$.

In the Table I the calculated masses of the $S$-wave baryons are shown \cite{17}.
In the first version we use only three parameters: the subenergy cutoff
$\lambda$ and the vertex function $g_0$, $g_1$, which corresponds to the
quark-quark interaction in $0^+$ and $1^+$ states. In this case the mass
values of strange baryons with $J^P=\frac{1}{2}^+$ are less than the
experimental ones. This means that the contribution color-magnetic interaction is too
large. In the second version we introduce four parameters: cutoff
$\lambda_0$, $\lambda_1$ and the vertex function $g_0$, $g_1$. We decrease
the color-magnetic interaction in $0^+$ strange channels and calculated mass
values of two baryonic multiplets  $J^P=\frac{1}{2}^+$, $\frac{3}{2}^+$
are in good agreement with the experimental data \cite{8}.

The essential difference between $\Sigma$ and $\Lambda$ is the spin of the
lighter diquark. The model explains both the sign and magnitude of this mass
splitting.

The suggested method of the approximate solution of the relativistic
three-quark equations allows us to calculate the $S$-wave baryons mass
spectrum. The interaction constants, determined the baryons spectrum in
our model, are similar to ones in the bootstrap quark model of $S$-wave
mesons \cite{15}. The diquark interaction forces are defined by the gluon
exchange. The relative contribution of the instanton-induced interaction is
less than that with the gluon exchange. This is the consequence of
$1/N_c$-expansion \cite{15}.

The gluon exchange corresponds to the color-magnetic interaction, which
is responsible for the spin-spin splitting in the hadron models. The sign
of the color-magnetic term is such as to made any baryon of spin
$\frac{3}{2}$ heavier than its spin-$\frac{1}{2}$ counterpart (containing
the same flavours).

\begin{table}
\caption{Baryon masses $M(J^p)$ (GeV).
Version 1 (the cutoff parameter $\lambda_1=12.2$), version 2 ($\lambda_0=9.7$,
$\lambda_1=12.2$). The vertex functions $g_0=0.702$, $g_1=0.540$. Experimental
values of the baryon masses \cite{8} are given in the parentheses.
\label{ta1}}
\begin{tabular}{ccp{4cm}ccc}
\toprule[1pt]
 & $M(\frac{1}{2}^+)$ &  & & $M(\frac{3}{2}^+)$ &
\\[5pt]
\midrule[1pt]
$N$ &
\begin{tabular}{c}
$0.940$ \\
$0.940$
\end{tabular}
& $(0.940)$ & $\Delta$ &
\begin{tabular}{c}
$1.232$ \\
$1.232$
\end{tabular}
& $(1.232)$
\\[5pt]
$\Lambda$ &
\begin{tabular}{c}
$1.022$ \\
$1.098$
\end{tabular}
& $(1.116)$ & $\Sigma^*$ &
\begin{tabular}{c}
$1.377$ \\
$1.377$
\end{tabular}
& $(1.385)$
\\[5pt]
$\Sigma$ &
\begin{tabular}{c}
$1.050$ \\
$1.193$
\end{tabular}
& $(1.193)$ & $\Xi^*$ &
\begin{tabular}{c}
$1.524$ \\
$1.524$
\end{tabular}
& $(1.530)$
\\[5pt]
$\Xi$ &
\begin{tabular}{c}
$1.162$ \\
$1.325$
\end{tabular}
& $(1.315)$ & $\Omega$ &
\begin{tabular}{c}
$1.672$ \\
$1.672$
\end{tabular}
& $(1.672)$
\\[5pt]
\bottomrule[1pt]
\end{tabular}
\end{table}

\section{Five-quark amplitudes for the charmed pentaquarks}\label{sec3}

In the following, we introduce how to get the relativistic five-quark amplitudes for the charmed pentaquarks, where we adopt the dispersion
relation technique. Due to the rules of $1/N_c$ expansion \cite{18, 19, 20}, we only need to consider planar diagrams, while the other diagrams can be neglected.
By summing over all possible subamplitudes which correspond to the division of complete system into subsystems smaller number of particles, we can obtain
the total amplitude.

In general, a five-particle amplitude ($\mathcal{A}$) can be expressed as the sum of ten subamplitudes ($\mathcal{A}_{ij}$ ($i=1,2,3,4$, $j=1,2,3,4,5$)), i.e.,

\begin{eqnarray}
\nonumber
\mathcal{A}&=&\mathcal{A}_{12}+\mathcal{A}_{13}+\mathcal{A}_{14}+\mathcal{A}_{15}+\mathcal{A}_{23}
+\mathcal{A}_{24}\nonumber\\&&+\mathcal{A}_{25}+\mathcal{A}_{34}+\mathcal{A}_{35}+\mathcal{A}_{45}, \nonumber
\end{eqnarray}

\noindent
where $\mathcal{A}_{ij}$ denotes the subamplitude from the pair interaction of particles $i$ and $j$ in a five-particle system.

For the sake of simplifying the calculation, we take the relativistic generalization of the Faddeev-Yakubovsky approach. With the $uuuu\bar{c}$ system as
an example, we introduce how to obtain $\mathcal{A}_{12}$.  Firstly, we need to construct the five-quark amplitude of the $uuuu\bar{c}$ system, where only
pair interaction with the quantum numbers of a $J^P=1^+$ diquark is included. Then,
the set of diagrams relevant to the amplitude $\mathcal{A}_{12}$ can further be broken down into groups corresponding to
amplitudes: $A_1(s,s_{1234},s_{12},s_{34})$, $A_2(s,s_{1234},s_{25},s_{34})$, $A_3(s,s_{1234},s_{13},s_{134})$, $A_4(s,s_{1234},s_{24},s_{234})$, which
are shown in Fig. 3 by the graphic representation of the equations for the five-quark subamplitudes. Similarly, we also give the corresponding graphic
representation for the $uuud\bar{c}$ and $udud\bar{c}$ systems, which are shown in Fig. 4.  For the cases of the $uuud\bar{c}$ and $udud\bar{c}$ systems,
there are six and seven subamplitudes, respectively. Here, the coefficients can be obtained by the permutation of quarks \cite{21, 22}.

In the following, we need to further illustrate how to write out the subamplitudes $A_1(s,s_{1234},s_{12},s_{34})$, $A_2(s,s_{1234},s_{25},s_{34})$,
$A_3(s,s_{1234},s_{13},s_{134})$, and $A_4(s,s_{1234},s_{24},s_{234})$, which are in the form of a dispersion relation. Firstly, we need to define the
amplitudes of quark-quark and quark-antiquark interaction $a_n(s_{ik})$. With the help of four-fermion interaction with quantum numbers of the
gluon \cite{15}, we can calculate the amplitudes $q\bar{q}\to q\bar{q}$ and $qq\to qq$ through the dispersion $N/D$ method. Thus, the pair quarks
amplitude can be expressed as \cite{15}

\begin{eqnarray}
a_n(s_{ik})&=&\frac{G_n^2(s_{ik})}{1-B_n(s_{ik})},\nonumber\\
B_n(s_{ik})&=&\int_{(m_1+m_2)^2}^{\Lambda_n}
\frac{ds_{ik}'}{\pi}\frac{\rho_n(s_{ik}')G_n^2(s_{ik}')}{s_{ik}'-s_{ik}}, \nonumber
\end{eqnarray}

\noindent
where $s_{ik}$ denotes the two-particle subenergy squared. And $s_{ijk}$ is the energy squared of particles $i$, $j$, $k$ while $s_{ijkl}$ is the
four-particle subenergy squared. In addition, we also define $s$ as the system total energy squared.

We obtain the concrete forms of $A_{i}$ ($i=1,2,3,4$), i.e.,

\begin{eqnarray}
&&A_1(s,s_{1234},s_{12},s_{34})\nonumber\\&&=\frac{\lambda_1B_3(s_{12})B_2(s_{34})}
{[1-B_3(s_{12})][1-B_2(s_{34})]}+6\hat{J}_2(3,2)
A_4(s,s_{1234},s_{23}',s_{234}')  \nonumber\\
&&\quad+2\hat{J}_2(3,2)A_3(s,s_{1234},s_{13}',s_{134}') +2\hat{J}_1(3)A_3(s,s_{1234},s_{15}',s_{125})\nonumber\\
&&\quad+2\hat{J}_1(2)A_4(s,s_{1234},s_{25}',s_{125})+4\hat{J}_1(2)A_4(s,s_{1234},s_{35}',s_{345}),\nonumber\\\label{k1}
\end{eqnarray}
\begin{eqnarray}
&&A_2(s,s_{1234},s_{25},s_{34})\nonumber\\&&=\frac{\lambda_2B_2(s_{25})B_2(s_{34})}
{[1-B_2(s_{25})][1-B_2(s_{34})]}+12\hat{J}_2(2,2)
A_4(s,s_{1234},s_{23}',s_{234}') \nonumber\\
&&\quad+8\hat{J}_1(2)A_3(s,s_{1234},s_{25}',s_{125}), \label{k2}
\end{eqnarray}
\begin{eqnarray}
&&A_3(s,s_{1234},s_{13},s_{134})\nonumber\\&&=\frac{\lambda_3B_3(s_{12})}
{1-B_3(s_{12})}+12\hat{J}_3(3)A_1(s,s_{1234},s_{12}',s_{34}'), \label{k3}
\end{eqnarray}
\begin{eqnarray}
&&A_4(s,s_{1234},s_{24},s_{234})\nonumber\\&&=\frac{\lambda_4B_2(s_{24})}
{1-B_2(s_{24})}+4\hat{J}_3(2)A_2(s,s_{1234},s_{25}',s_{34}')\nonumber\\
&&\quad+4\hat{J}_3(2)A_1(s,s_{1234},s_{12}',s_{34}'),\label{k4}
\end{eqnarray}

\noindent
where $\lambda_i$ denotes the current constants. In addition, the  integral operators $\hat{J}_1(l)$, $\hat{J}_2(l,p)$, and $\hat{J}_3(l,p)$ are introduced,
where their expressions can be found in Appendix.
Taking the same treatment as that given in Ref. \cite{23}, where we pass from the integration over the cosines of the angles to the integration
over the subenergies, we can extract
two-particle singularities in the amplitudes $A_1(s,s_{1234},s_{12},s_{34})$, $A_2(s,s_{1234},s_{25},s_{34})$, $A_3(s,s_{1234},s_{13},s_{134})$, and
$A_4(s,s_{1234},s_{24},s_{234})$:

\begin{eqnarray*}
A_1(s,s_{1234},s_{12},s_{34})&=&\frac{\alpha_1
(s,s_{1234},s_{12},s_{34})B_3(s_{12})B_2(s_{34})}{[1-B_3(s_{12})]
[1-B_2(s_{34})]},
\end{eqnarray*}
\begin{eqnarray*}
A_2(s,s_{1234},s_{25},s_{34})&=&\frac{\alpha_2
(s,s_{1234},s_{25},s_{34})B_2(s_{25})B_2(s_{34})}{[1-B_2(s_{25})]
[1-B_2(s_{34})]},
\end{eqnarray*}
\begin{eqnarray*}
A_3(s,s_{1234},s_{13},s_{134})&=&\frac{\alpha_3
(s,s_{1234},s_{13},s_{134})B_3(s_{13})}{1-B_3(s_{13})},
\end{eqnarray*}
\begin{eqnarray*}
A_4(s,s_{1234},s_{24},s_{234})&=&\frac{\alpha_4
(s,s_{1234},s_{24},s_{234})B_2(s_{24})}{1-B_2(s_{24})}.
\end{eqnarray*}

\noindent
Here, we want to further specify that we do not extract the three-particle and four-particle singularities, which are weaker than the two-particle
singularities. In addition, we also adopt the classification of singularities suggested in Ref. \cite{16}.
The main singularities in $s_{ik}=(m_i +m_k)^2$ there are from pair rescattering of
particles $i$ and $k$. First of all, they are threshold square-root singularities. Also
possible are singularities which correspond to the bound states.
We have apart from two-particle singularities the triangular
singularities, the singularities defining the interaction of four and five particles. Such
classification allows us to search the corresponding solution by taking
into account some definite number of leading singularities and neglecting all the weaker
ones. We consider the approximation which defines two-particle, three-, four-, and
five-particle singularities.
As the smooth functions of $s_{ik}$, $s_{ijk}$, $s_{ijkl}$, and $s$,
$\alpha_1(s,s_{1234},s_{12},s_{34})$, $\alpha_2(s,s_{1234},s_{25},s_{34})$, $\alpha_3(s,s_{1234},s_{13},s_{134})$ and
$\alpha_4(s,s_{1234},s_{24},s_{234})$ can be expanded in a series in the singularity point, where only the first term of this series should be
employed further. Thus, we further define the reduced amplitudes $\alpha_1$, $\alpha_2$, $\alpha_3$, $\alpha_4$, and the B-functions in the middle
point of the physical region of Dalitz-plot at the point $s_0$, i.e.,

\begin{eqnarray}
s_0^{ik}&=&{s_0}=\frac{s+3\sum_{i=1}^5m_i^2}{0.25\sum_{i,k=1,i\neq k}^5(m_i+m_k)^2},
\end{eqnarray}
\begin{eqnarray}
s_{123}&=&0.25s_0\sum_{i,k=1,i\neq k}^3(m_i+m_k)^2-\sum_{i=1}^3m_i^2,
\end{eqnarray}
\begin{eqnarray}
s_{1234}&=&0.25s_0\sum_{i,k=1,i\neq k}^4(m_i+m_k)^2-2\sum_{i=1}^4m_i^2.
\end{eqnarray}
Then, we replace the integral Eqs.  (\ref{k1})-(\ref{k4}) corresponding to the diagrams in Fig. 3 by the following algebraic equations
\begin{eqnarray}
\alpha_1&=&\lambda_1+6\alpha_4J_2(3,2,2)+2\alpha_3J_2(3,2,3)
+2\alpha_3J_1(3,3)\nonumber\\
&&+2\alpha_4J_1(3,2)+4\alpha_4J_1(2,2),\label{m1}\\
\alpha_2&=&\lambda_2+12\alpha_4J_2(2,2,2)+8\alpha_3J_1(2,3),\\
\alpha_3&=&\lambda_3+12\alpha_1J_3(3,3,2),\\
\alpha_4&=&\lambda_4+4\alpha_2J_3(2,2,2)+4\alpha_1J_3(2,2,3),\label{m2}
\end{eqnarray}
respectively. Here, the definitions of the functions $J_1(l,p)$, $J_2(l,p,r)$, $J_3(l,p,r)$ ($l,p,r$= 1, 2, 3) are listed in Appendix.

Finally, we have the function like
\begin{eqnarray}
\alpha_i(s)=F_i(s,\lambda_i)/D(s),
\end{eqnarray}
where the masses of these discussed systems can be determined by zeros of $D(s)$ determinants. And, $F_i(s,\lambda_i)$ denotes the function
of $s$ and $\lambda_i$, which determines the contribution of subamplitude.

\section{Numerical results}\label{sec4}

In Sec. \ref{sec3}, the involved parameters in our model include quark masses $m_{u,d}=439$ MeV and $m_c=1640$ MeV, where we effectively take into account
the contribution of the confinement potential in obtaining the spectrum of charmed pentaquarks. The adopted value of cutoff $\Lambda=10$, which coincides
with that taken in Ref. \cite{24, 25}. In addition, a dimensionless parameter $g$, which is the gluon coupling constant, is
introduced in our calculation. We notice that the mass of charmed pentaquark with both configuration $D_2^* N$ ($udud\bar{c}$) and quantum number
$(I)J^P=(0)\frac{3}{2}^+$ was calculated through the one-boson-exchange model  in Ref. \cite{26}, where its mass is $3387$ MeV. Thus, by
reproducing this value in our model, we can determine $g=0.825$, which is adopted in the following calculation to give more predictions of the masses of
charmed pentaquarks.

With the above preparation, in this section we present the numerical results of the mass spectrum of the discussed charmed pentaquarks, where the poles
of the reduced amplitudes $\alpha_1$, $\alpha_2$, $\alpha_3$, $\alpha_4$ correspond to the bound states of charmed pentaquarks. The predicted masses of
charmed pentaquarks are shown in Table \ref{ta2}.

\renewcommand{\arraystretch}{1.6}
\begin{table}[htbp]
\caption{The obtained low-lying charmed pentaquark masses. Here, the parameters involved in our model include: quark mass $m_{u,d}$ = 439 MeV, $m_c$ = 1640 MeV;
cutoff parameter $\lambda$ = 10; and gluon coupling constant $g$ = 0.825. Here, -- denotes that there does not exist the corresponding charmed pentaquark
state. \label{ta2}}
\begin{center}
\begin{tabular}{ccccccc} \toprule[1pt]
&\multicolumn{4}{c}{$J^P$}\\
States&$\frac{1}{2}^+$&$\frac{3}{2}^+$&$\frac{1}{2}^-$&$\frac{3}{2}^-$\\\midrule[1pt]
$\Theta_c^{++}(uuuu\bar{c})/\Theta_c^{--}(dddd\bar{c})$&3323&3323&3339&3339\\
$\Theta_c^+(uuud\bar{c})/\Theta_c^-(dddu\bar{c})$&2986&3209&3277&--\\
$\Theta_c^0(udud\bar{c})$&2980&3387&3280&--\\
\bottomrule[1pt]
\end{tabular}
\end{center}
\end{table}

\section{Summary}\label{sec5}

As an interesting research reach topic, exploring the exotic multiquark matter beyond conventional meson and baryon is an exciting and important task, which
will be helpful to understand the non-perturbative behavior of quantum chromodynamics. The new observation of numerous $XYZ$ particles opens the Pandora's Box
of studying the exotic multiquark matter \cite{2}.

In this work, we studied the charmed pentaquarks with $J^P=1/2^\pm,3/2^\pm$ by the relativistic five-quark model, where the Faddeev-Yakubovsky type
approach is adopted. The masses of the low-lying charmed pentaquarks are calculated. This information is useful to further experimental search for them in future.

We used some approximations for the calculation of the five-quark amplitude.
The estimation of the theoretical error on the pentaquark masses is
about 20\%. It is usual for model estimations. This result was obtained by the choice of model parameters:
gluon coupling constant $g=0.825$ and cutoff parameter $\lambda=10$.

We also notice that there were several experimental efforts on the search for the charmed pentaquarks \cite{27, 28, 29}, where the present experiment still
did not find the evidence of charmed pentaquark. Unlike the mesons, all half-integral spin and parity quantum numbers are allowed in the baryon sector, which
means that there exists the mixing between charmed pentaquark and conventional charmed baryon, so that experimentally searching for such charmed pentaquark
is not a simple task. In addition, the charmed pentaquarks have the abnormally small widths since the observed charmed pentaquarks with the isospin $I=0,1,2$
and the spin-parity $J^P=\frac{1}{2}^+$, $\frac{3}{2}^+$ are below the thresholds. These facts make the identification of a pentaquark be difficult in experiment.

In summary, exploring the charmed pentaquark is a reach field full of challenges and opportunities. More theoretical and experimental united efforts
should be made in the future to establish charmed exotic pentaquark family.

\section*{Acknowledgments}
This work was carried with the support of the RFBR, Research Project (Grant No. 13-02-91154). This project is also supported by the National Natural Science
Foundation of China under Grants No. 11222547, No. 11175073, No. 11035006 and No. 11311120054, the Ministry of Education of China
(SRFDP under Grant No.
2012021111000), the Fok Ying Tung Education Foundation
(Grant No. 131006).

\section*{Appendix: Some useful formulae}

The definitions of $\hat{J}_1(l)$, $\hat{J}_2(l,p)$ and $\hat{J}_3(l,p)$
are given by
\begin{eqnarray}
&&\hat{J}_1(l)\nonumber\\&&=\frac{G_l(s_{12})}{[1-B_l(s_{12})]}\int_{(m_1+m_2)^2}
^{\Lambda_l}\frac{ds_{12}'}{\pi}\frac{G_l(s_{12}')\rho_l(s_{12}')}
{s_{12}'-s_{12}}\int_{-1}^{+1}\frac{dz_1}{2},\nonumber\\\label{h1}
\end{eqnarray}
\begin{eqnarray}
&&\hat{J}_2(l,p)\nonumber\\&&=\frac{G_l(s_{12})G_p(s_{34})}{[1-B_l(s_{12})]
[1-B_p(s_{34})]}\int_{(m_1+m_2)^2}^{\Lambda_l}
\frac{ds_{12}'}{\pi} \nonumber \\
&&\quad\times\frac{G_l(s_{12}')\rho_l(s_{12}')}
{s_{12}'-s_{12}}\int_{(m_3+m_4)^2}^{\Lambda_p}
\frac{ds_{34}'}{\pi}\frac{G_p(s_{34}')\rho_p(s_{34}')}
{s_{34}'-s_{34}}\nonumber\\
&&\quad\times\int_{-1}^{+1}\frac{dz_3}{2}\int_{-1}^{+1}\frac{dz_4}{2},\label{h2}
\end{eqnarray}
\begin{eqnarray}
&&\hat{J}_3(l)\nonumber\\&&=\frac{G_l(s_{12},\widetilde{\Lambda})}
{1-B_l(s_{12},\widetilde{\Lambda})}\times\frac{1}{4\pi}
\int_{(m_1+m_2)^2}^{\widetilde{\Lambda}}
\frac{ds_{12}'}{\pi}\frac{G_l(s_{12}',\widetilde{\Lambda})\rho_l(s_{12}')}
{s_{12}'-s_{12}}\nonumber\\
&&\quad\times\int_{-1}^{+1}\frac{dz_1}{2}\int_{-1}^{+1}dz
\int_{-1}^{+1}dz_2 \nonumber \\
&&\quad\times\frac{1}{\sqrt{1-z^2-z_1^2-z_2^2+2zz_1z_2}},\label{h3}
\end{eqnarray}
respectively, where $l,p$ are taken as 1, 2, 3. And $m_i$ denotes the corresponding quark mass. In Eqs. (\ref{h1}) and (\ref{h3}), $z_1$ is the cosine of
the angle between the relative momentum of the particles 1 and 2 in the intermediate state and the momentum of the particle 3 in the final state, taken
in the c.m. of particles 1 and 2. In Eq. (\ref{h3}), we can define $z$ as the cosine of the angle between the momenta of the particles 3 and 4 in the final
state, taken in the c.m. of particles 1 and 2. $z_2$ is the cosine of the angle between the relative momentum of particles 1 and 2 in the intermediate state
and the momentum of the particle 4 in the final state, is taken in the c.m. of particles 1 and 2. In  Eq. (\ref{h2}), $z_3$ is the cosine of the angle
between relative momentum of particles 1 and 2 in the intermediate state and the relative momentum of particles 3 and 4 in the intermediate state, taken
in the c.m. of particles 1 and 2. $z_4$ is the cosine of the angle between the relative momentum of the particles 3 and 4 in the intermediate state and
that of the momentum of the particle 1 in the intermediate state, taken in the c.m. of particles 3 and 4.

In Eqs. (27) -- (29),
 $G_n(s_{ik})$ denote the quark-quark and quark-antiquark vertex functions, where the concrete expressions of $G_n(s_{ik})$ are listed in
Table. \ref{ta3}. The vertex functions satisfy the Fierz relations. All of these vertex functions are generated from $g_V$, $g_V^{(c)}$.
Dimensional constants of the four-fermion interaction $g_V$, $g_V^{(c)}$ are parameters of model. In order to avoid additional violation
parameters we introduce the scale of the dimensional parameters similar to (6). Dimensionless parameters $g$ and $\Lambda$ are supposed
to be constants which independent of the quark interaction type.

\renewcommand{\arraystretch}{1.6}
\begin{table}[htb]
\caption{The expressions of vertex function $G_n(s_{ik})$. \label{ta3}}
\begin{center}
\begin{tabular}{cc} \toprule[1pt]
$J^{PC}$     &$G_n^2$  \\ \midrule[1pt]           
$0^+ (n=1)$  &$4g/3-2g(m_i+m_k)^2/(3s_{ik})$ \\
$1^+ (n=2)$  &$2g/3$         \\
$0^{++} (n=3)$  &$8g/3$      \\
$0^{-+} (n=4)$  &$8g/3-4g(m_i+m_k)^2/(3s_{ik})$ \\
\bottomrule[1pt]
\end{tabular}
\end{center}
\end{table}

In Eqs. (27) -- (29), $B_n(s_{ik})$ is the Chew-Mandelstam function, where $\Lambda_n$ is the cutoff  \cite{11}. Additionally, we also list
the expression of the phase space $\rho_n(s_{ik})$, i.e.,
\begin{eqnarray}
\rho_n(s_{ik},J^{PC})&=&\bigg(\alpha(J^{PC},n)\frac{s_{ik}}{(m_i+m_k)^2}+
\beta(J^{PC},n) \nonumber \\
&&+\gamma(J^{PC},n)\frac{(m_i-m_k)^2}{s_{ik}}\bigg) \\
&&\times\frac{\sqrt{[s_{ik}-(m_i+m_k)^2][s_{ik}-(m_i-m_k)^2]}}{s_{ik}}, \nonumber
\end{eqnarray}
where the values of $\alpha(J^{PC},n)$, $\beta(J^{PC},n)$, and $\gamma(J^{PC},n)$ are shown in Table \ref{ta4}.

\renewcommand{\arraystretch}{1.6}
\begin{table}[htb]
\caption{The coefficients of Chew-Mandelstam functions. Here,
$n=1$ corresponds to a $qq$ -pair with $J^P=0^+$ in the $\bar{3}_c$ color state, while $n=2$ describes a $qq$ -pair with $J^P=1^+$ in the $\bar{3}_c$ color
state. And then, $n=3$ defines the $q\bar q$-pairs corresponding to mesons with quantum numbers $J^{PC}=0^{++}$ and $0^{-+}$. In addition, we also
define $e=(m_i-m_k)^2/(m_i+m_k)^2$. \label{ta4}}
\begin{center}
\begin{tabular}{ccccccc} \toprule[1pt]
$J^{PC}$     &~~$n$~~  &$\alpha(J^{PC})$ &$\beta(J^{PC})$ &$\gamma(J^{PC})$  \\ \midrule[1pt]           
$0^{++}$     &3  &$1/2$      &$-1/2$   &0  \\
$0^{-+}$     &3  &$1/2$      &$-e/2$   &0  \\
$0^{+}$      &1  &$1/2$      &$-e/2$   &0  \\
$1^{+}$      &2  &$1/3$      &$1/6-e/3$ &$-1/6$  \\
\bottomrule[1pt]
\end{tabular}
\end{center}
\end{table}

In addition, we also list the definitions of some functions used in this work, i.e.,
\begin{eqnarray}
&&J_1(l,p)\nonumber\\&&=\frac{{G_l^2(s_{0}^{12})}B_p(s_0^{13})}{B_l(s_0^{12})}\int_{(m_1+m_2)^2}
^{\Lambda_l}\frac{ds_{12}'}{\pi}\frac{\rho_l(s_{12}')}
{s_{12}'-s_0^{12}}\nonumber\\
&&\quad\times\int_{-1}^{+1}\frac{dz_1}{2}\frac{1}{1-B_p(s_{13}')},
\end{eqnarray}
\begin{eqnarray}
&&J_2(l,p,r)\nonumber\\&&=\frac{G_l^2(s_0^{12})G_p^2(s_0^{34})B_r(s_0^{13})}
{B_l(s_0^{12})B_p(s_0^{34})}\times\int_{(m_1+m_2)^2}^{\Lambda_l}
\frac{ds_{12}'}{\pi}\frac{\rho_l(s_{12}')}
{s_{12}'-s_0^{12}}\nonumber\\
&&\quad\times{\int_{(m_3+m_4)^2}^{\Lambda_p}}
\frac{ds_{34}'}{\pi}\frac{\rho_p(s_{34}')}
{s_{34}'-s_0^{34}}
\int_{-1}^{+1}\frac{dz_3}{2}\nonumber\\
&&\quad\times\int_{-1}^{+1}\frac{dz_4}{2}
\frac{1}{1-B_r(s_{13}')},
\end{eqnarray}
\begin{eqnarray}
&&J_3(l,p,r)\nonumber\\&&=\frac{G_l^2(s_0^{12},\widetilde{\Lambda})
B_p(s_0^{13})B_r(s_0^{24})}
{1-B_l(s_0^{12},\widetilde{\Lambda})}\frac{1-B_l(s_0^{12})}
{B_l(s_0^{12})}\nonumber\\
&&\quad\times\frac{1}{4\pi}
\int_{(m_1+m_2)^2}^{\widetilde{\Lambda}}
\frac{ds_{12}'}{\pi}\frac{\rho_l(s_{12}')}
{s_{12}'-s_0^{12}}\int_{-1}^{+1}\frac{dz_1}{2}\nonumber\\
&&\quad\times\int_{-1}^{+1}dz
{\int_{z_2^-}^{z_2^+}}dz_2\frac{1}{\sqrt{1-z^2-z_1^2-z_2^2+2zz_1z_2}}\nonumber\\
&&\quad\times
\frac{1}{[1-B_p(s_{13}')][1-B_r(s_{24}')]},
\end{eqnarray}
Since other choices of point $s_0$ do not change essentially the contributions of $\alpha_1$, $\alpha_2$, $\alpha_3$ and $\alpha_4$, the
indexes $s_0^{ik}$ are omitted here. Due to the weak dependence of the vertex functions on the energy, we treat them as constants in our calculation,
which is an approximation. The details of the integration contours of the function {$J_1,J_2,J_3$} can be found in Ref. \cite{30}.

\begin{figure*}[htbp]
\includegraphics[bb=250 10 330 700 ,scale=0.3]{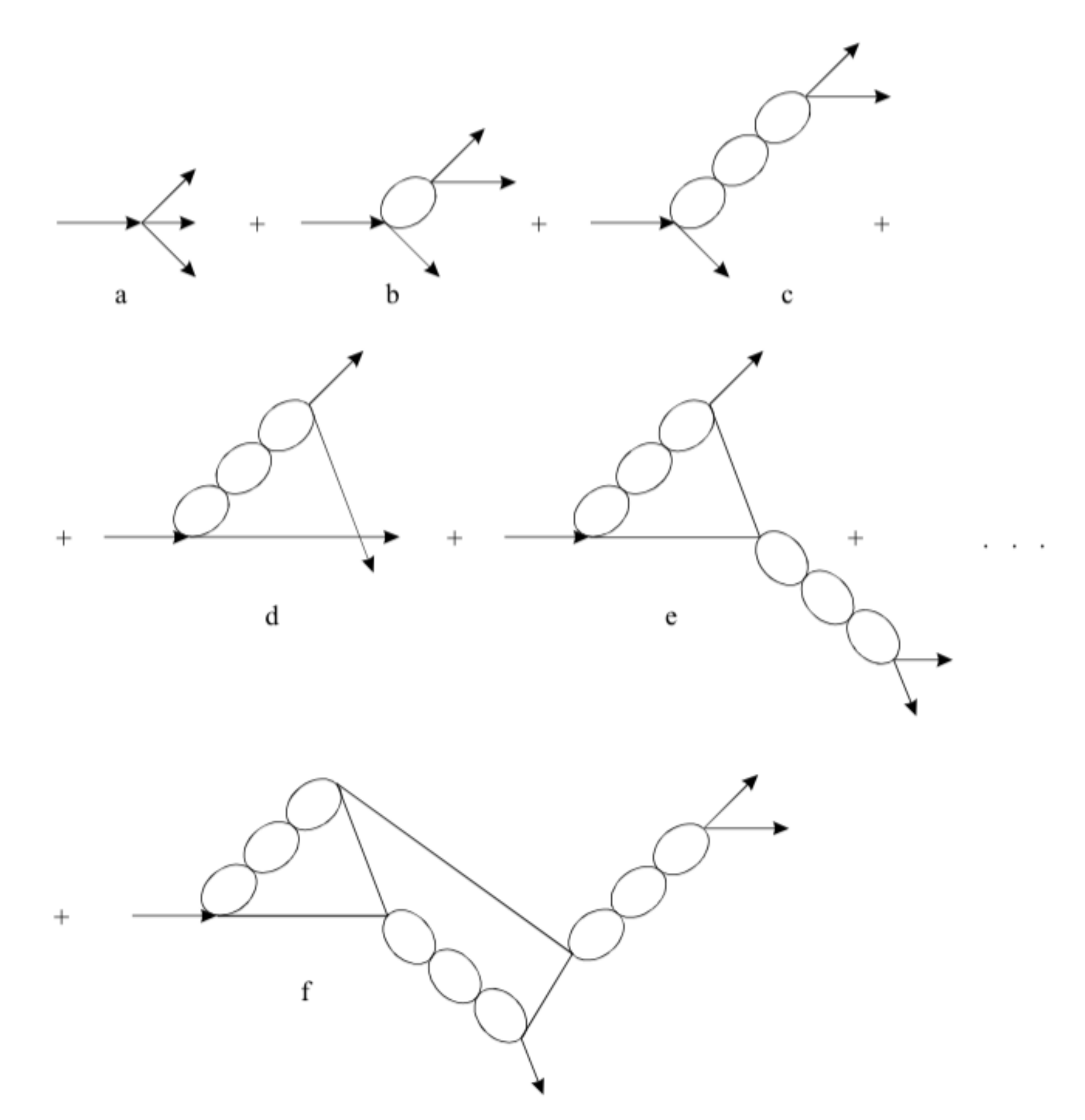}
\caption{Diagrams which correspond to a) production of three quarks, b-f) subsequent pair interaction.\label{fig3}}
\end{figure*}

\begin{figure*}[htbp]
\includegraphics[bb=250 0 480 0 ,scale=0.5]{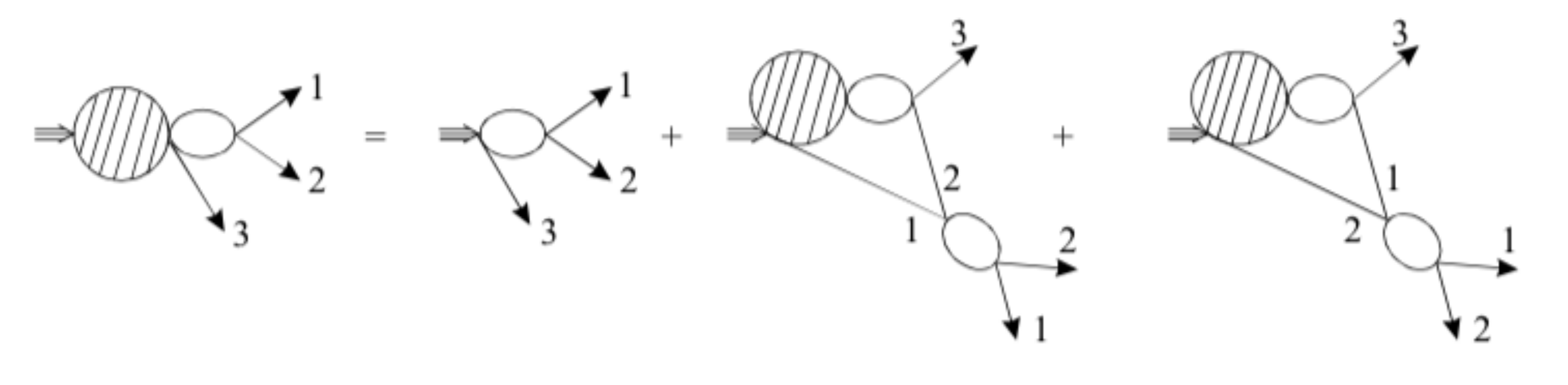}
\caption{Graphic representation of the equation for the amplitude $A_1(s,s_{12})$ (formulae (7) and (11)).\label{fig4}}
\end{figure*}

\begin{figure*}[htbp]
\includegraphics[bb=250 10 330 100 ,scale=0.69]{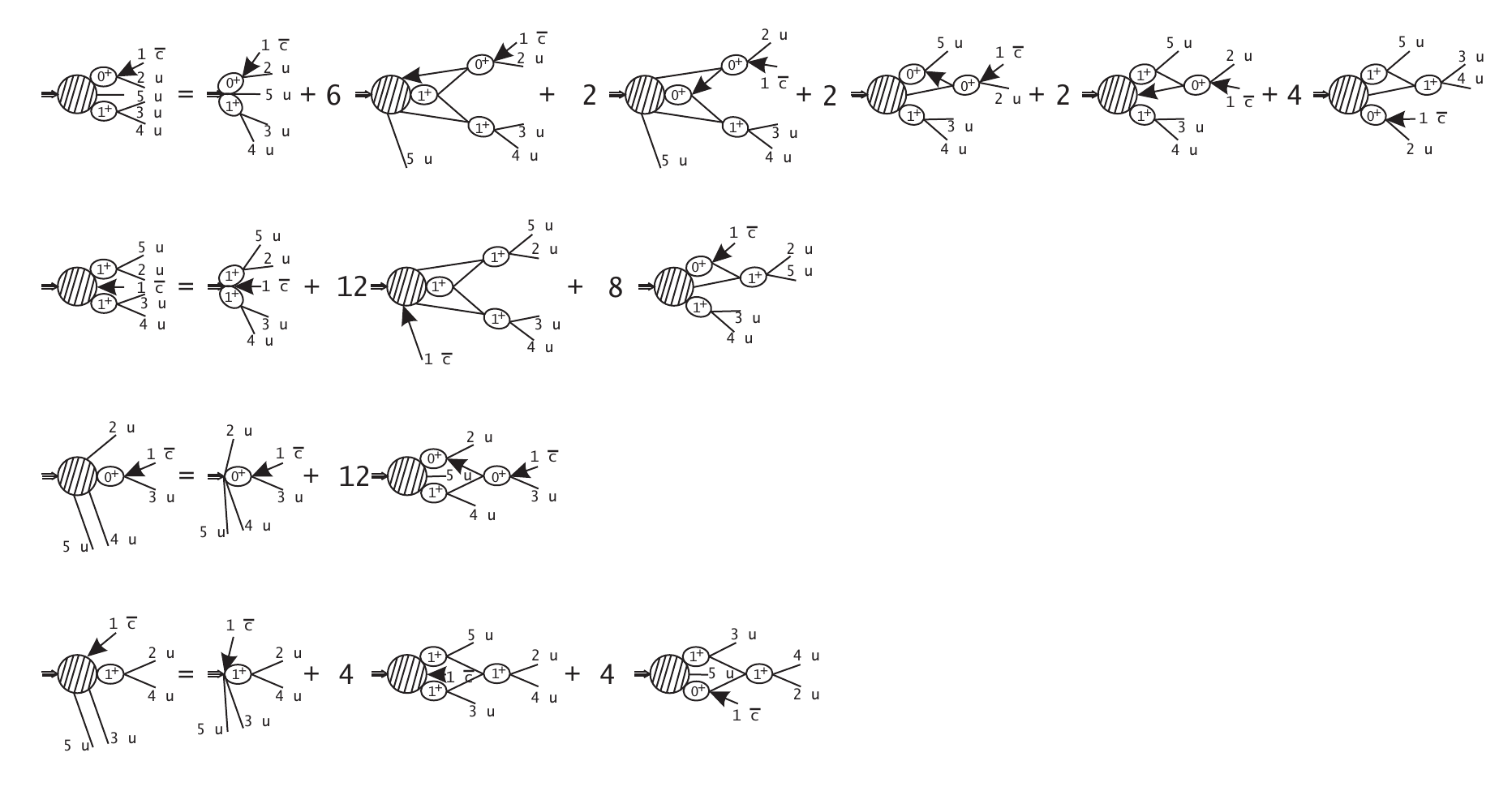}
\caption{The graphic representation of the equations for the five-quark subamplitudes $A_k$ ($k$=1-4) in the case of the $uuuu\bar{c}$ system. Here, we mark $\bar{c}$ and other four light quarks by lines with and without arrow, respectively. \label{fig1}}
\end{figure*}

\begin{figure*}[htbp]
\includegraphics[bb=250 10 330 320 ,scale=0.69]{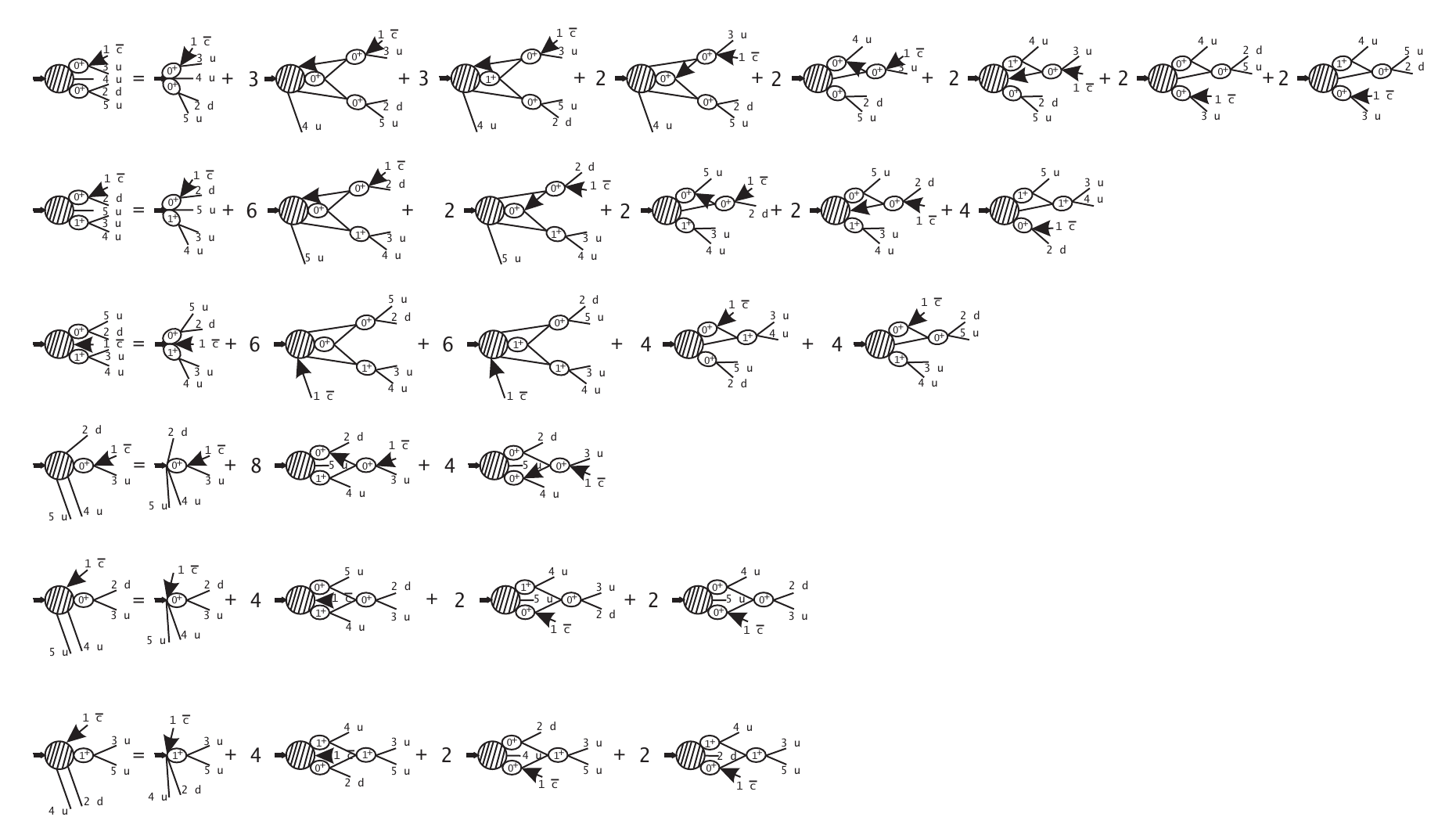}\\ 
\includegraphics[bb=250 10 330 320 ,scale=0.69]{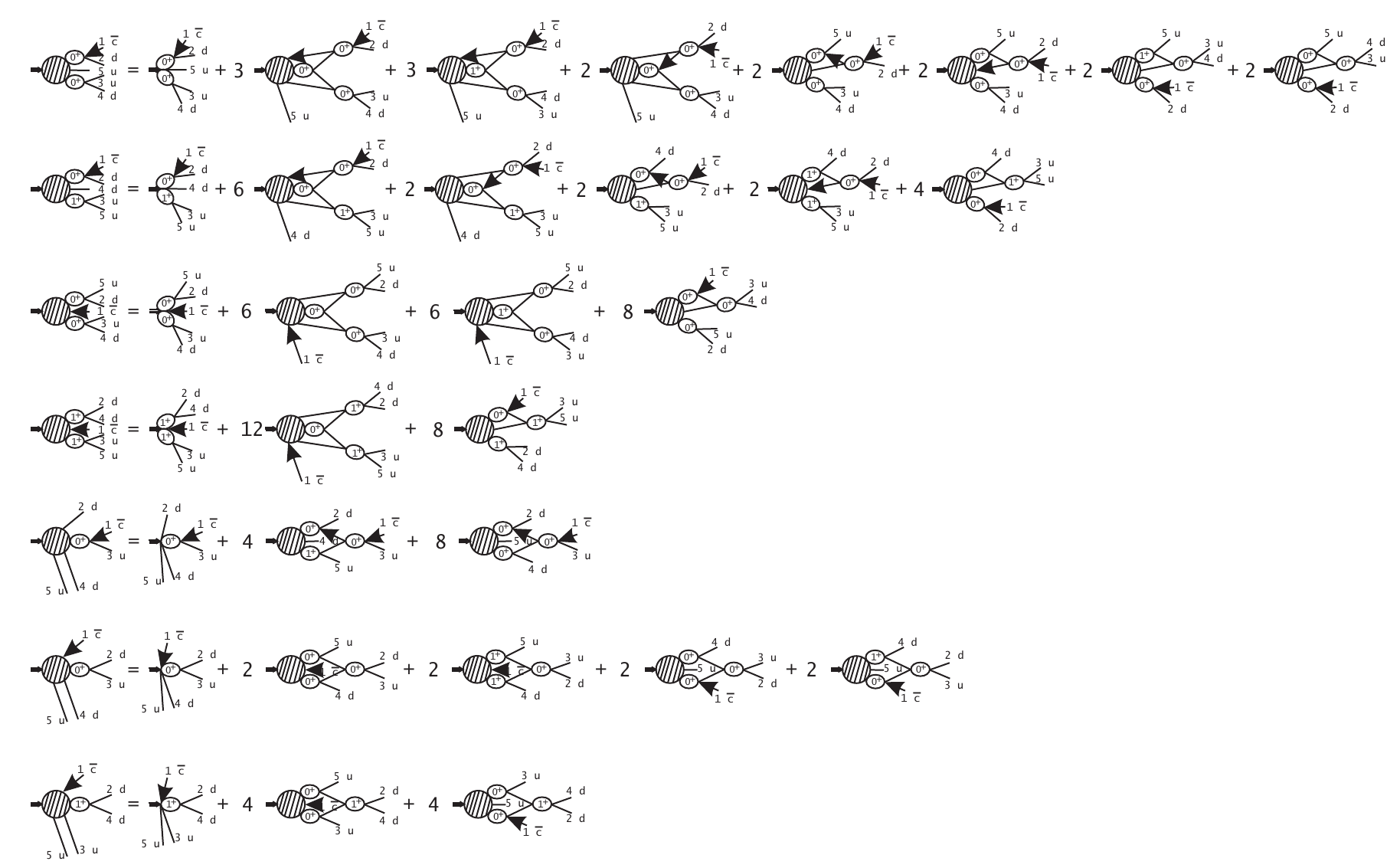}
\caption{The graphic representation of the equations for the five-quark subamplitudes for the $uuud\bar{c}$ and $udud\bar{c}$ systems. Here, the $\bar{c}$ quark is denoted by the lines with arrow.  There are six and seven diagrams for the $uuud\bar{c}$ and $udud\bar{c}$ systems, respectively.\label{fig2}}
\end{figure*}

\end{document}